**Comment** on "Case for a U(1)$_\pi$ Quantum Spin Liquid Ground State in the Dipole-Octupole Pyrochlore Ce$_2$Zr$_2$O$_7$" by E. M. Smith *et al*., Phys. Rev. X **12**, 021015 (2022).

-------------------------------------------------------------------------------------------------------------------

S. W. Lovesey[1,2]

[1]ISIS Facility, STFC, Didcot, Oxfordshire OX11 0QX, UK

[2]Diamond Light Source, Harwell Science and Innovation Campus, Didcot, Oxfordshire OX11 0DE, UK

**Abstract** All interpretations of extensive magnetic neutron scattering data are at fault. For the expressions for scattered intensities Eqs. (E1), (E2) and (F1) do not contain the cerium multipole whose likely contribution to the magnetic state of the pyrochlore is a principal goal of the published investigation. The Comment includes a brief look at essential corrections to the theory.

-------------------------------------------------------------------------------------------------------------------

Van Vleck in 1939 addressed the use of neutron scattering to investigate exchange forces in a paramagnetic material and chose MnS as an example [1]. More recently, Smith *et al*. [2] used the same experimental technique in a study of a pyrochlore material hosting tripositive cerium ions with essentially the same objective. In addition to exchange forces, Ref. [2] reports a search for evidence of so-called dipole-octupole Ce states. To meet these ends, Smith *et al*. confront extensive sets of measured neutron scattering intensities with simulated intensities. The latter are not meaningful, however, because pseudospin operators and not true Ce tensor operators are used. The sole place of pseudospin-1/2 operators in describing properties of Ce$_2$Zr$_2$O$_7$ is one of labelling Ce$^{3+}$ (4f$^1$) electronic degrees of freedom.

De Gennes [3] advanced Van Vleck's study and Brockhouse [4] surveyed some early investigations of neutron scattering. Specifically, in 1938, Halpern and Johnson [5] proved that intensities diminish on moving away from the forward direction of scattering because of an atomic form factor. Intensities Eqs. (E1), (E2) and (F1) in Ref. [2] are not attended by an explicit definition of the atomic form factor common to all three expressions. However, it is without doubt inadequate, because dipole and octupole contributions possess different form factors; the dipole form factor discussed by Halpern and Johnson [5] is finite in the forward direction of scattering whereas the octupole form factor is zero in that position [6-10]. Likely, Smith *et al*. [2] were snared into using a dipole atomic form factor by the erroneous use of pseudospin (dipole) operators. At best, the corresponding simulations illustrate intensities near the forward direction where contributions from octupoles are entirely negligible.

True spin **s**, and linear momentum **p** and a conjugate position **R** are electronic degrees of freedom illuminated by a beam of neutrons. Matrix elements of compound spin [exp($i\boldsymbol{\kappa} \cdot \mathbf{R}$) **s**] and orbital [exp($i\boldsymbol{\kappa} \cdot \mathbf{R}$) **e** × **p**] operators appear in a magnetic scattering amplitude **Q**$_\perp$, where **e** is a unit vector in the direction of the reflection vector $\boldsymbol{\kappa}$. Inclusion of orbital matrix elements in **Q**$_\perp$ are essential for rare earth ions with significant orbital angular momentum [6, 7]. Thereafter, matrix elements in **Q**$_\perp$ are replaced by standards methods in favour of equivalent matrix elements of magnetic spherical tensors. These are axial or polar tensors depending on whether an ion is at a centre of inversion symmetry [8-13]. Cerium ions occupy centrosymmetric sites ($D_{3d}$, $\bar{3}m$) in space group $Fd\bar{3}m$ (16(c) in No. 227). A measurement of the crystal field potential of Ce in $Ce_2Zr_2O_7$ shows that a Kramers doublet is formed by a state $|u\rangle$ and its conjugate $|\bar{u}\rangle$ with,

$$|u\rangle = a|J, M\rangle + b|J, -M\rangle. \qquad (1)$$

Atomic states from the configuration $^2F$ in Eq. (1) have a total angular momentum $J = 5/2$ and a projection $M = 3/2$ [2, 14]. A ground state $|g\rangle = [|u\rangle + i|\bar{u}\rangle]/\sqrt{2}$ obeys site symmetry $D_{3d}$ for real coefficients in $|u\rangle$. Intensity of total scattering measured by Smith *et al*. [2] is derived with $\langle g|\mathbf{Q}_\perp \cdot \mathbf{Q}_\perp|g\rangle$ attended by the spatial distribution of Ce ions, and such a simulation is beyond the scope of this Comment.

Expectation values of $\mathbf{Q}_\perp$ decompose the Kramers state set in a specific point group, a decomposition that mimics the work of pseudospins. Axial magnetic multipoles are labelled $t^K_Q$ or $T^K_Q$ for one of two point-groups used here for illustration. The integer K is the multipole rank and $(2K + 1)$ projections obey $-K \leq Q \leq K$. Boothroyd provides a tutorial guide to the expectation value of $\mathbf{Q}_\perp$ for $Ce^{3+}$ in $CePd_2Si_2$, and it is highly anisotropic in this material, i.e., a strong function of $\mathbf{e}$ [15].

Let us consider axial multipoles $t^K_Q$ prescribed by site symmetry $D_{3d}$ ($\bar{3}m$). Radial integrals in $Ce^{3+}$ form factors depend on the magnitude of the reflection vector $\kappa$, and they are denoted by $\langle j_m(\kappa)\rangle$ with m = 2, 4 and 6, and $\langle j_m(0)\rangle = 0$ [16]. One finds,

$$\langle t^3_{+3}\rangle = \langle g|t^3_{+3}|g\rangle = i(4/7)\sqrt{(1/35)}\,h(\kappa),\ \langle t^5_{+3}\rangle = -i(5/11)\sqrt{(1/231)}\,g(\kappa), \qquad (2)$$

where $h(\kappa) = \{\langle j_2(\kappa)\rangle + (10/3)\langle j_4(\kappa)\rangle\}$ and $g(\kappa) = \{\langle j_4(\kappa)\rangle + 12\langle j_6(\kappa)\rangle\}$ are octupole and the triakontadipole form factors, respectively, which vanish in the forward direction of scattering. Results for $\langle t^K_Q\rangle$ are independent of the specific values of coefficients a and b in $|u\rangle$, i.e., multipoles in question are symmetry-protected.

Symmetry $D_{3d}$ must be modified to yield a magnetic dipole, and point-group $\bar{3}m'$ qualifies to this end. The corresponding dipole $\langle T^1_0\rangle$ is aligned with the triad axis of rotation symmetry in space-group $Fd\bar{3}m'$ (No. 227.131, BNS). The centrosymmetric crystal class $m\bar{3}m'$ ($T_h$) allows a piezomagnetic effect, while it is not compatible with ferromagnetism and the only magnetoelectric effect is non-linear HEE (H & E are magnetic and electric fields, respectively). From Eq. (1),

$$\langle T^1_0\rangle = \langle u|T^1_0|u\rangle = (1/3)\{\mu\,\langle j_0(\kappa)\rangle + [\langle L_0\rangle + (12/35)(a^2 - b^2)]\,\langle j_2(\kappa)\rangle\}. \qquad (3)$$

Here, $\langle j_0(0)\rangle = 1$ and the form factor used in Ref. [2] is possibly modelled on $\langle j_0(\kappa)\rangle$. The magnetic moment $\mu = (6/7)\langle J_0\rangle = (9/7)(a^2 - b^2)$ and orbital angular momentum $\langle L_0\rangle = (4/3)\mu$. Higher-order multipoles in the scattering amplitude allowed by $\bar{3}m'$ are,

$$\langle T^3{}_0 \rangle = -\mu\,(14/45)\,\sqrt{(1/7)}\,h(\kappa), \quad \langle T^3{}_{+3} \rangle = -ab\,(8/7)\,\sqrt{(1/35)}\,h(\kappa),$$

$$\langle T^5{}_0 \rangle = -\mu\,(5/99)\,\sqrt{(5/33)}\,g(\kappa), \quad \langle T^5{}_{+3} \rangle = ab\,(10/11)\,\sqrt{(1/231)}\,g(\kappa). \qquad (4)$$

Multipoles with an even rank are absent because Eq. (1) contains one J-state [16]. Such multipoles represent entanglement of anapole and spatial degrees of freedom.

To summarize, Smith *et al*. [2] simulate observable quantities - scattering intensities - using pseudo-operators that are irrelevant in the material world. In consequence, the authors mine no sound evidence for or against the $U(1)_\pi$ state in $Ce_2Zr_2O_7$ from their extensive neutron scattering intensities. The powder average of the diffraction intensity is a plausible guide to a total intensity $\langle \mathbf{Q}_\perp \cdot \mathbf{Q}_\perp \rangle$, i.e., $\langle \mathbf{Q}_\perp \cdot \mathbf{Q}_\perp \rangle \approx \{\langle \mathbf{Q}_\perp \rangle \cdot \langle \mathbf{Q}_\perp \rangle\}_{av}$ [16]. Here, multipoles with ranks K = 1 (dipole), 3 (octupole), and 5 (triakontadipole) in the expectation value of the scattering amplitude $\langle \mathbf{Q}_\perp \rangle$ are calculated with the Ce Kramers doublet derived from a measured crystal field potential [2, 14]. Likewise, for spin-flip and non-spin-flip intensities to confront with intensities extracted by neutron polarization analysis [2, 17].

**Acknowledgement**. Dr D. D. Khalyavin (ISIS) provides ongoing support and guidance to my research. Professor A. T. Boothroyd (Oxford University) kindly perused the Comment in its making.

[1] J. H. Van Vleck, Phys. Rev. **55**, 924 (1939).

[2] E. M. Smith *et al*., Phys. Rev. X **12**, 021015 (2022).

[3] P. G. De Gennes, J. Phys. Chem. Solids **4**, 223 (1958).

[4] B. N. Brockhouse, Rev. Mod. Phys. **67**, 735 (1995).

[5] O. Halpern and M. H. Johnson, Phys. Rev. **55**, 818 (1938).

[6] G. T. Trammell, Phys. Rev. **92**, 1387 (1953).

[7] D. F. Johnston, Proc. Phys. Soc. **88**, 37 (1966).

[8] S. W. Lovesey, E. Balcar, K. S. Knight, and J. F. Rodriguez, Phys. Rep. **411**, 233 (2005).

[9] S. W. Lovesey, Phys. Scr. **90**, 108011(2015).

[10] M.-T. Suzuki, H. Ikeda, and P. M. Oppeneer, J. Phys. Soc. Jpn. **87**, 041008 (2018).

[11] K. W. H. Stevens, Proc. Phys. Soc. A **65**, 209 (1952).

[12] R. J. Elliott and K. W. H. Stevens, Proc. Roy. Soc. A **218**, 553 (1953).

[13] B. R. Judd, *Operator Techniques in Atomic Spectroscopy* (McGraw-Hill, New York,1963).

[14] R. Sibille *et al*., Nat. Phys. **16**, 546 (2020).

[15] A. T. Boothroyd, *Principles of Neutron Scattering from Condensed Matter* (OUP, Oxford UK, 2020).

[16] S. W. Lovesey and G. van der Laan, Phys. Rev. B **101**, 144419 (2020).

[17] S. W. Lovesey, Phys. Rev. B **106**, 064415 (2022).